\documentclass[pra,twocolumn,showpacs,amsmath,amssymb,eqsecnum,nofootinbib,floatfix]{revtex4}
\usepackage{epsfig}
\usepackage{bm} 
\usepackage{amsmath}

\def\cf{{\it cf.~}}
\def\erf{\text{erf}}
\begin{document}

\title{Representations of  Spacetime Alternatives and Their Classical Limits}

\author{August W.~Bosse}
\email{awbosse@physics.ucsb.edu}
\affiliation{Department of Physics, University of California\\
Santa Barbara, CA 93106-9530 USA} 

\author{James B.~Hartle}
\email{hartle@physics.ucsb.edu}
\affiliation{Department of Physics, University of California\\
Santa Barbara, CA 93106-9530 USA}

\date{\today}

\begin{abstract}

Different quantum mechanical operators can correspond to the same classical
quantity.  Hermitian operators differing only by operator ordering of the
canonical coordinates and momenta at one moment of time are the most
familiar example. Classical spacetime alternatives that extend over
time can also be represented by different quantum operators. For example,
operators representing a particular value of the time average of a
dynamical variable can be constructed in two ways: First, as the projection
onto the value of the time averaged Heisenberg picture operator for the
dynamical variable.  Second, as the class operator defined by a
sum over those histories of the dynamical variable that have the specified
time-averaged value.  We show both by explicit example and general argument that
the predictions of these different representations agree in the classical limit and that sets of histories represented by them decohere in that limit. 

\end{abstract}

\maketitle

\section{Introduction}

Usual quantum mechanics predicts the probabilities of alternatives 
specified at a moment in time and histories of such alternatives specified 
at a sequence of times. A single particle moving in one dimension provides 
a familiar example. The probability $p(\Delta)$ that the particle's 
position $x$ lies in a range $\Delta$ at a time $t$ is given in the 
Heisenberg picture by
	\begin{equation}
	p(\Delta) = \|
	P^x_{\Delta}(t) | \psi\rangle \|^{2}\, .
	\label{eq:prob}
	\end{equation}
Here, $P^x_{\Delta}(t)$ is the projection onto the range $\Delta$ of the 
eigenvalues of the operator $x$ a time $t$, and $| \psi \rangle$ is the 
particle's state.

A given classical alternative, for example $2x^3p^2$ (where $p$ is the 
momentum), can correspond to several distinct quantum mechanical 
alternatives, e.g., $x^3 p^2 + p^2x^3$, $xp^2x^2 + x^2p^2x$, $xpxpx+ xpxpx$.
These differ by operator ordering. Probabilities of suitably coarse grained
ranges of such alternatives approximately agree for states representing classical 
situations. This paper explores different operator 
representations of a more general type of alternative for a single
non-relativistic particle --- spacetime alternatives extended over time.

Classical alternatives are not restricted to definite moments of time. 
Consider the continuous average of position over the range of 
times between $0$ and $T$, specifically
	\begin{equation}
	\overline{x} \equiv \frac{1}{T}\int_{0}^{T} x(t) \, dt\, .
	\label{eq:ave2}
	\end{equation}
This is a simple example of a spacetime alternative that is not at one 
time, but rather is extended over time.

The general notion of spacetime alternatives for a single particle moving 
in  one dimension is  a partition of the set of paths of the particle into 
an exhaustive set of mutually exclusive classes. For example, we could 
partition the paths into classes defined by whether the values of a functional, such as  
Eq.~(\ref{eq:ave2}), fall into one or another of an exhaustive set of ranges 
$\{\Delta_{\alpha}\}$, $\alpha = 1,\,2\,...$. We could partition the paths by 
whether or not they cross a given spacetime region between two times, 
etc. Alternatives at a single moment of time are just a special case of this more 
general class.

Spacetime alternatives in the sense of field averages occur routinely in 
field theory \cite{Haa96}. Spacetime alternatives may permit more realistic 
descriptions of measurements. No realistic measurement occurs
exactly at one moment in time. Finally, spacetime alternatives may be
essential for a quantum theory of gravity, where there is no definite
notion of spacetime geometry to supply meaning to
``at a moment of time'' \cite{Har95c}.

How are spacetime alternatives represented in quantum mechanics? The 
consideration of spacetime alternatives in quantum mechanics has a long 
history. The discussions in \cite{BR33,BR50,Fey48,DeW62,Men93,Har91b, 
YT91a, Sor93, Mar94,Halxx,Har04} are just some of the many examples that 
could be cited. In the context of non-relativistic quantum mechanics, a 
comprehensive treatment can be given in the sum-over-histories quantum 
mechanics of closed systems \cite{Har91b,YT91a}. The essential feature used in this 
paper is the following:  If the paths $x(t)$ between $t=0$ and $t=T$ are 
partitioned into an exhaustive set of mutually exclusive classes 
$c_{\alpha}$,  $\alpha = 1,\,2\,...$, then the operators 
$\hat{C}_{\alpha}$ representing the individual alternatives in this set of
histories are defined in the Schr\"odinger picture\footnote{We use a hat to distinguish the class
operator defined by the path integral \eqref{onethree} in the
Schr\"odinger picture from the sums of chains of projections $C_\alpha$
useful the Heisenberg picture. They are related by
$\hat C_\alpha = \exp(-iHT/\hbar) C_\alpha$. 
Both operators give the same probabilities in \eqref{eq:classprob}.} 
by sums over the histories
in $c_{\alpha}$. Specifically these sums have the  form
	\begin{equation}
	\langle x'' | \hat{C}_{\alpha} | x' \rangle \equiv 
	\int_{c_{\alpha}} \delta x \,
	\exp{\left( \frac{i}{\hbar}S[x(t)] \right) }\, . 
	\label{onethree}
	\end{equation}
Here, $S[x(t)]$ is the action functional. The sum is over all paths 
$x(t)$ which start at $x'$ at $t=0$, end at $x''$ at $t=T$, and are in 
the class $c_{\alpha}$. As an example, consider the set of histories defined by whether
the time average $\bar x$ in Eq.\eqref{eq:ave2} lies  in one
or another of an exhaustive set of exclusive ranges $\{\Delta_\alpha\}, \alpha= 1,2 \cdots$.
The path integral that defines the 
class operator
$\hat{C}_{\Delta}$ for the alternative that 
$\overline{x}$ lies in the particular
range $\Delta$ in the  the set $\{\Delta_\alpha\}$ is over all paths
with the above starting and ending values for which $\overline{x}$ has
a value in $\Delta$. Its probability is
	\begin{equation}
	p_{soh}(\Delta) = \| \hat{C}_{\Delta} | \psi \rangle
	\|^{2} \ .
	\label{eq:classprob}
	\end{equation}
        
The set of probabilities defined by Eq.\eqref{eq:classprob} for all $\Delta$ in the 
set  $\{\Delta_\alpha\}$  are generally not consistent with the rules of probability
theory unless a decoherence condition is satisfied \cite{Har91b,YT91a}. For instance, the probability
for $\bar x$ to lie in a range $\Delta$ and the probability to lie in the complementary
range $\bar\Delta$ would not generally sum to one. An example of a condition which
ensures that such relations are satisfied is the medium decoherence condition.
	\begin{equation}
	\langle \psi | {\hat C}^\dagger_\Delta {\hat C}_{\Delta'} | \psi \rangle \approx 0,
	\quad \Delta \ne \Delta' . 
	\label{meddec}
	\end{equation}
When this is satisfied for all set pairs $\Delta\ne\Delta'$ drawn from the $\{\Delta_\alpha\}$ the set of histories is said to {\it decohere.} 

The time average $\overline{x}$ also defines a Hermitian operator in the 
Heisenberg picture. It is therefore also natural to think of its 
probability as being given by
	\begin{equation}	
	p_{proj}(\Delta) = \| {P}_{\Delta} | \psi\rangle \|^{2}
	\label{eq:projprob}
	\end{equation}
where ${P}_{\Delta}$ is the projection operator onto the 
eigenstates of the operator $\overline{x}$ defined by Eq.\eqref{eq:ave2}. Decoherence is not an issue for the alternatives defined by the set projections onto the ranges $\{\Delta_\alpha\}$.
The analog of the decoherence condition Eq.\eqref{meddec} is automatically
satisfied because projections onto different ranges are exactly orthogonal:
	\begin{equation}
	P_\Delta P_{\Delta'} = 0, \quad \Delta \ne \Delta' .
	\label{orthproj}
	\end{equation}

Eq. \eqref{eq:projprob} is the rule 
usually given for the probabilities of measurements of fields averaged 
over spacetime regions \cite{Haa96}. There are questions as to what 
such measurements might mean \cite{Mar94}, when their outcomes are accurately 
predicted by Eq.~(\ref{eq:projprob}), and how to assign probabilities to 
sequences of such measurements\cite{Sor93}. This paper leaves such interesting issues 
aside and instead concentrates  on analyzing the mathematical 
difference between Eq.~(\ref{eq:classprob}) and Eq.~(\ref{eq:projprob}) 
in simple models, asking whether their predictions coincide in the classical limit, and determining whether or not the alternatives $\hat C_\alpha$ decohere in that limit.

At the level of operators, it is convenient to compare $\hat
C_\Delta$,  not with $P_\Delta$ directly, but rather with the 
combination
	\begin{equation}
	\hat P_\Delta \equiv e^{-iHT/\hbar} P_\Delta 
	\label{eq:Pdeltahat}
	\end{equation}
which is not a projection but  gives the same probabilities as Eq.~\eqref{eq:projprob} and satisfies [cf. Eq.\eqref{orthproj}]
\begin{equation}
\hat P^\dagger_\Delta \hat P_{\Delta'} = 0, \quad \Delta \ne \Delta' .
	\label{orthproj1j}
	\end{equation}
This comparison is useful because $\hat P_\Delta$ and $\hat C_\Delta$ coincide in the limit of large widths of $\Delta$. 

In Sections IV and V we calculate the action of  
$\hat{C}_{\Delta}$ and $\hat{P}_{\Delta}$ on simple wave functions 
for two models where they can be evaluated explicitly---a
non-relativistic free particle and a non-relativistic harmonic oscillator. 
We show that the two operators 
are quantitatively nearly the same when acting on states describing
classical situations. Further we show that they coincide in a formal $\hbar\to 0$ 
limit. Section VI gives a general argument why this is the case, and Section VII    shows
how sets of such alternatives decohere in the $\hbar\to 0$ limit.   We conclude that
$\hat{C}_{\Delta}$ and 
$\hat{P}_{\Delta}$ can be regarded as distinct 
quantum mechanical representations of the same classical alternative.

\section{Formalism}
\setcounter{equation}{0}
In this section we will briefly outline how to construct the 
quantum operators $\hat C_\Delta$ and $\hat P_\Delta$ defined in 
Eqs.~\eqref{onethree} and \eqref{eq:Pdeltahat} that  
represent the spacetime alternative $\overline{x} \in \Delta$ defined in 
Eq.~\eqref{eq:ave2}. 

The coarse grained history of interest $c_{\Delta}$ is the class of
histories $x(t)$ defined as follows
	\begin{equation}
	c_{\Delta} = 
	\left\{ x(t) \left|\overline{x} \equiv \frac{1}{T} \int_{0}^{T} dt \, x(t) \in \Delta \right. \right\}
	\label{eq:avex}
	\end{equation}
where $\Delta$ is the subset of the real line of width $\delta$ centered on
$x_{c}$:
	\begin{equation}
	\Delta = \left\{ x | x_{c} - \delta/2 \leq x \leq
	x_{c} + \delta/2 \right\} \ .
	\label{twotwo}
	\end{equation} 
That is, the class $c_{\Delta}$ consists of
all paths that start at $x'$ at $t=0$ and 
end at $x''$ at $t=T$, such that the average value of the path 
$\overline{x}$ is in the range $\Delta$. The operator $\hat C_\Delta$ is
defined by the path integral in Eq.~\eqref{onethree} over paths in the
class $c_\Delta$. 
The operator $\hat P_\Delta$ is defined by:
	\begin{align}
	\langle x'' | \hat{P}_{\Delta} | x' \rangle =&
	\langle x'' | e^{-iHT/\hbar} P_{\Delta} | x' \rangle  \nonumber  \\
	 = &
	\int_{\Delta}   d\overline{x} \, 
	\langle x''| e^{-iHT/\hbar} | \overline{x} \rangle \langle 
	\overline{x} | x' \rangle\, .
	\end{align}

How do these two operators differ?  
Clearly $\hat{C}_{\Delta}$ and 
$\hat{P}_{\Delta}$ are not equal. $\hat{P}_{\Delta}$ is proportional to 
a single projection operator, while $\hat{C}_{\Delta}$ is proportional
to an infinite product of projection operators which in general is not a 
projection operator.  However, for large $\delta$,  $\hat{C}_{\Delta}$ and $\hat{P}_{\Delta}$ 
approach each other. To see this, let us examine  how 
$\hat{C}_{\Delta}$ and $\hat{P}_{\Delta}$ are constructed in more detail.

The eigenstates of $\overline{x}$ form a complete set with delta function 
normalization:
	\begin{equation}
	\int_{-\infty}^{\infty} d\overline{x}
	\, | \overline{x} \rangle \langle \overline{x} 
	| = I ,
	\; \;
	\langle	\overline{x} | \overline{x}' \rangle =
	\delta ( \overline{x} - \overline{x}' )\, .
	\label{eq:complete}
	\end{equation}
The projection, ${P}_{\Delta}$ is defined by:
	\begin{equation}
	P_{\Delta} = \int_{\Delta} d\overline{x} \, | \overline{x} \rangle
	\langle \overline{x} |\, .
	\label{eq:defP}
	\end{equation}
As $\delta$ increases, the integral in Eq.~(\ref{eq:defP}) 
is over a larger and larger portion of the real line defined by
Eq.~\eqref{twotwo}. Thus, from Eq.~(\ref{eq:complete}), 
as $\delta$ approaches infinity, $P_{\Delta}$ 
approaches unity, and  
	\begin{align}
	\lim_{\delta \rightarrow \infty}& \langle x''| e^{-iHT/\hbar} 
	P_{\Delta} | x' \rangle \nonumber \\ & =
	\langle x'' |  e^{-iHT/\hbar} | x' \rangle \equiv K(x'',T;x',0)
	\end{align}
where $K(x'',T;x',0)$ is the propagator from $t=0$ to $t=T$.

For $\langle x'' |\hat{C}_{\Delta}|x'\rangle$ we are integrating over all 
paths between $x'$ at $t=0$ and $x''$ at $t=T$, such that $\overline{x} \in 
\Delta$. Thus, as $\delta$ approaches infinity, the class of  
paths being integrated over becomes less and less restricted, and 
	\begin{equation}
	\lim_{\delta \rightarrow \infty} \int_{c_{\Delta}} \delta x \,
	e^{iS[x(t)]/\hbar} = \int _u \delta x \,
        e^{iS[x(t)]/\hbar} = K(x'',T;x',0)
	\label{twoseven}
	\end{equation}
where the unrestricted  functional integral on the right hand side is over {\it all}  paths 
between $x'$ at $t=0$ and $x''$ at $t=T$. Therefore, in the limit of
large $\delta$, we find
	\begin{equation}
	\langle x'' | \hat{P}_{\Delta} | \psi \rangle \approx
	\langle x'' | \hat{C}_{\Delta} | \psi \rangle\, .
	\end{equation}
for suitable initial  states $|\psi\rangle$. 
The scale of $\delta$ above which this approximate equality holds is set by the 
spatial extent of the wave function $\psi(x,t)$ over the time interval $t \in (0,\,T)$.
 
In subsequent sections we evaluate and compare  $\langle x'' |
\hat{P}_{\Delta} | \psi \rangle$ and  
$\langle x'' | \hat{C}_{\Delta} | \psi \rangle$ explicitly using 
Gaussian initial wave functions $\psi(x)$ for  
two simple systems: the one-dimensional, non-relativistic free 
particle, and the one-dimensional, non-relativistic harmonic oscillator. 

\section{General Potential}
\setcounter{equation}{0}

Consider a one-dimensional quantum system with the Hamiltonian
	\begin{equation}
	H=\frac{p^{2}}{2m}+V(x)\, .
	\label{threeone}
	\end{equation}
In Section IV we will set $V(x) = 0$, and in Section V
we will set $V(x) = \frac{1}{2}m\omega^{2}x^{2}$, but here we discuss
results that do not depend on these specific forms of $H$. 

In the Heisenberg picture, an operator $O$ evolves in time according to the 
equation of motion
	\begin{equation}
	i\hbar\frac{dO}{dt} = [O,\,H]\, .
	\label{threetwo}
	\end{equation}
For $p$ and $x$, this gives the following coupled evolution equations
	\begin{equation}
	\dot{x}(t) = \frac{p}{m} \ ,  \qquad
	\dot{p}(t) = -\frac{dV(x)}{dx}\, .
	\end{equation}
With a suitable operator ordering prescription,  these equations can be 
solved for $x(t)$ and $p(t)$ in terms of $p(0) \equiv  p_{0}$ and 
$x(0)\equiv x_{0}$. We then construct $\overline{x}$ from the 
Heisenberg picture operator $x(t)$:
	\begin{equation}
	\overline{x} = \frac{1}{T} \int_{0}^{T} dt \, x(t)\, .
	\label{threefour}
	\end{equation}
The eigenstates $|\overline{x}\rangle$ of $\overline{x}$ form a complete 
set with delta function normalization, Eq.~(\ref{eq:complete}). One 
can construct an operator representation of the alternative
$\overline{x} \in \Delta$ by projecting onto the eigenvalues of 
$\overline{x}$, as in Eq.~(\ref{eq:defP}). For the time period starting at 
$t=0$ and ending at $t=T$, this operator has matrix elements in the 
position basis given by
        \begin{eqnarray}
        \langle x'' | \hat{P}_{\Delta} | x' \rangle
	&=&
	\langle x'' | e^{-iHT/\hbar} P_{\Delta} | x' \rangle
	 \label{eq:pop} \\
        &=& \int_{\Delta} d\overline{x} \int_{-\infty}^{\infty} dy 
	\, \langle x'' |
	e^{-iHT/\hbar} | y \rangle \langle y | \overline{x} \rangle 
	\langle \overline{x} |
	x' \rangle\, .
	\nonumber
	\end{eqnarray}

Calculating the class operator $\hat{C}_{\Delta}$ for the spacetime 
alternative $\overline{x} \in \Delta$ is slightly less straightforward. We 
begin with $\hat{C}_{\Delta}$ defined in \eqref{onethree}.
The functional integral can be rewritten  by introducing the top-hat
function
	\begin{equation}
	e_{\Delta}(z) = \
	\left\{
	\begin{array}{l}
	1, \; z\in \Delta \\
	0, \; z \not\in \Delta
	\end{array} 
	\right.
	\label{eq:char}
	\end{equation}
and its Fourier transform
	\begin{equation}
	e_{\Delta}(z) = \frac{1}{\sqrt{2\pi}} \int_{-\infty}^{\infty} dk 
	\, e^{ikz}
	\tilde{e}_{\Delta}(k)\, .
	\end{equation}
Thus we may write Eq.~(\ref{onethree}) as
	\begin{eqnarray}
	 \langle x'' | \hat{C}_{\Delta} | x' \rangle
         &\equiv&\int_{c_{\Delta}} \delta x \, e^{iS[x(t)]/\hbar} \nonumber \\ 
        &=&
	\int_u \delta x \, e_{\Delta}(\overline{x}[x(t)]) e^{iS[x(t)]/\hbar} 
	\label{eq:classe} \\
	&=& \frac{1}{\sqrt{2\pi}} \int_{-\infty}^{\infty} dk 
	\, \tilde{e}_{\Delta}(k)
	\int_u \delta x \, e^{iS[x(t)]/\hbar + ik \overline{x}}
        \nonumber
	\end{eqnarray}
where the unrestricted path integral is now over all paths from $x'$ at $t=0$ to $x''$
at $t=T$ and $\overline x[x(t)]$ is the time averaging functional defined by
Eq.~\eqref{threefour}. Eq.~(\ref{eq:classe}) allows us to define an effective action
        \begin{equation}
        S_{\rm{eff}}[x(t)] = S[x(t)] + \hbar k \overline{x}
        \label{effaction}
        \end{equation}
and an effective Lagrangian
        \begin{equation}
        L_{\rm{eff}} = L + \frac{\hbar}{T} k x(t)\, .
        \label{efflagrang}
	\end{equation}

We can take this calculation a step further by evaluating 
$\tilde{e}_{\Delta}(k)$. Note that $e_{\Delta}(z)=\int_{\Delta} dy \, 
\delta(z-y)$. Thus
	$$
	\tilde{e}_{\Delta}(k) = 
	\frac{2}{\sqrt{2\pi}} \frac{1}{k} e^{-ikx_{c}}
	\sin \left(\frac{k\delta}{2}\right)
	$$
and the class operator is thus given by
	\begin{equation}
	\langle x'' | \hat{C}_{\Delta} | x' \rangle = \frac{1}{\pi}
	\int_{-\infty}^{\infty} \frac{dk}{k} \,
	e^{-ikx_{c}}\sin{\left( \frac{k\delta}{2} \right) }
	\int_u \delta x \, e^{iS_{\rm{eff}}[x(t)]/\hbar}\, .
	\label{eq:classeff}
	\end{equation}

We are now prepared to examine these expressions for specific quantum 
systems. We discuss the free particle in Section IV and the harmonic 
oscillator in Section V.

\section{The Free Particle}
\setcounter{equation}{0}
\label{freepart}

\subsection{$\hat{P}_{\Delta}$ and $\hat{C}_{\Delta}$}
We begin by examining the free particle: $L = \frac{1}{2}m\dot{x}^{2}$. For 
this system the Heisenberg equations can be solved: 
$x(t)=x_{0}+(p_{0}/m)t$ and $p(t) = p_{0}$. Thus,
	\begin{equation}
	\overline{x} = x_{0} + \frac{p_{0}}{2m}T\, .
	\label{fourone}
	\end{equation}
In the position basis, the eigenstates of $\overline{x}$ are solutions of 
the equation
	\begin{equation}
	x \langle x | \overline{x} \rangle + \frac{T}{2m} \frac{\hbar}{i}
	\frac{d}{dx} \langle x | \overline{x} \rangle =
	\overline{x} \langle x | \overline{x} \rangle\, .
	\end{equation}
Solving this equation and imposing the delta function normalization, 
Eq.~(\ref{eq:complete}), gives:
	\begin{equation}
	\langle x | \overline{x} \rangle = \sqrt{\frac{m}{\pi \hbar T}}
	\exp{\left[\frac{i}{\hbar} \frac{2m}{T} \left(	
	\overline{x}x-\frac{x^{2}}{2}
	\right) \right] }\, .
	\label{eq:xbarpos}
	\end{equation} 
We substitute this expression into Eq.\eqref{eq:pop} along with the
expression for the propagator of the free particle:
	\begin{align}
	\langle x'' | e^{-iHT/\hbar} | x' \rangle &\equiv
	K(x'',T;x',0) \nonumber \\ 
	&= \sqrt{\frac{m}{2 \pi i \hbar T}}
	\exp{ \left[ \frac{i}{\hbar} \frac{m}{2T} (x''-x')^2 \right] }\, .
	\label{eq:freeprop}
	\end{align}
Integrating Eq.~(\ref{eq:pop}) over $\bar{x}$ and $y$ gives:
	\begin{equation}
	\langle x'' | \hat{P}_{\Delta} | x' \rangle
	=K(x'',T;x',0)
	E_\Delta\left(\frac{x''+x'}{2}, \lambda \right) .
	\label{eq:eqforP}
	\end{equation}
where $\Delta$ is the range $[x_c-\delta/2,x_c+\delta/2]$ and the length $\lambda$ is
defined by
	\begin{equation}
	\lambda \equiv \left( \frac{\hbar T}{2 m} \right)^{1/2}
	= 2.3 \times 10^{-14} \left(\frac {1 {\rm g}}{m}\right)^{1/2} \left(\frac{T}{1 {\rm s}}\right) {\rm cm} \ .
	\label{lambda}
	\end{equation}
Here, $E_\Delta(z,\ell)$ is the function on the range $\Delta=[a,b]$
defined by 
	\begin{equation}
	E_\Delta(z,\ell)\equiv \frac{1}{2}
	\left[\erf\left(\frac{z-a}{\sqrt{i}\ell}\right) -
	\erf\left(\frac{z-b}{\sqrt{i}\ell}\right)\right]
	\label{Edef}
	\end{equation}

\begin{figure}
	\centerline{\epsfxsize=3in, \epsfbox{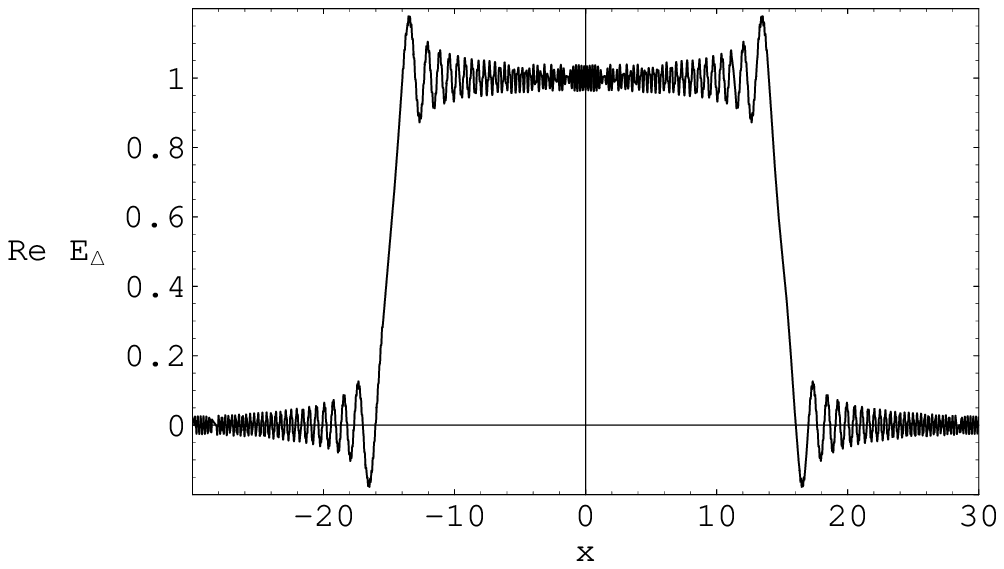}}
	\centerline{\epsfxsize=3in, \epsfbox{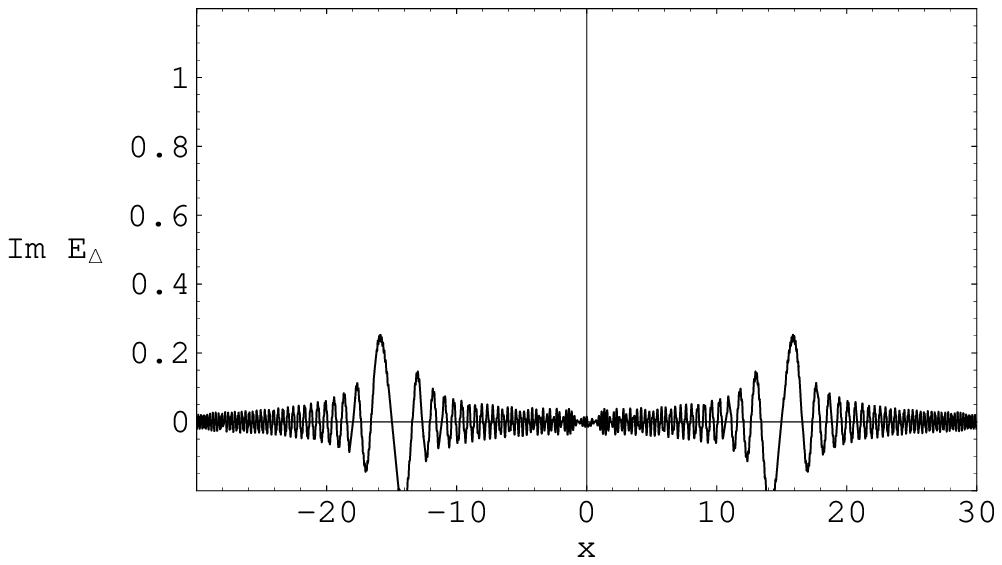}}
	\caption{The real and imaginary parts of the function 
	$E_\Delta(z,\ell)$ defined by \eqref{Edef}. These graphs are for 
	$\delta/\lambda =15$. Already at this value the	real part is a reasonable 
	approximation to a top hat function $e_\Delta(z/\ell$) and the imaginary 
	part is small. As $\hbar$ approaches zero, the value of $\lambda$ becomes very 
	small [cf Eq. \eqref{lambda}], the approximation of the real part to a top-hat function
	in both the expression Eq.~\eqref{eq:eqforP} for $\langle x'' | \hat{P}_{\Delta} | x' \rangle$ and
	Eq.~\eqref{eq:eqforC} for $\langle x'' | \hat{C}_{\Delta} | x' \rangle$ becomes
	better and better and the imaginary part becomes increasingly negligible.}
	\label{Efunction}
	\end{figure}

Figure \ref{Efunction} illustrates the function $E_\Delta(z,\ell)$. Its 
important property for the subsequent discussion is that it approaches the 
top-hat function $e_\Delta$ as the dimensionless ratio $\delta/\ell$ becomes 
large. Figure \ref{Efunction} shows that
this is not a bad approximation when that ratio is only 15. 

	\begin{figure}
	\begin{center}
			\epsfxsize=.93\linewidth
			\epsffile{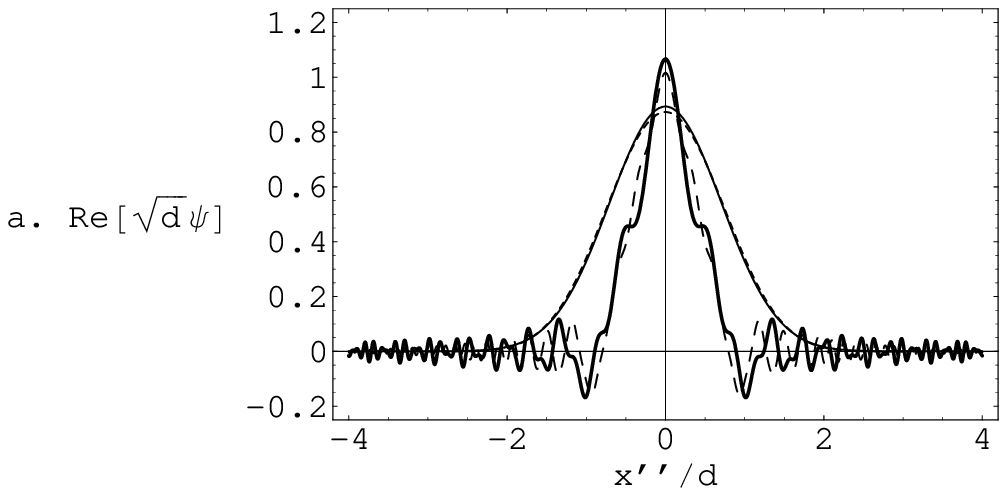}
			\epsfxsize=.93\linewidth
			\epsffile{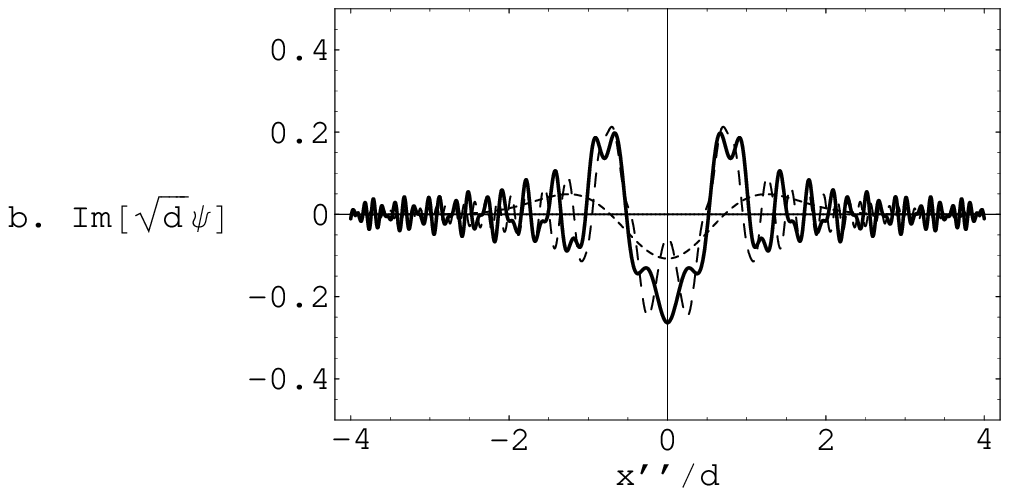}
			\epsfxsize=.93\linewidth
			\epsffile{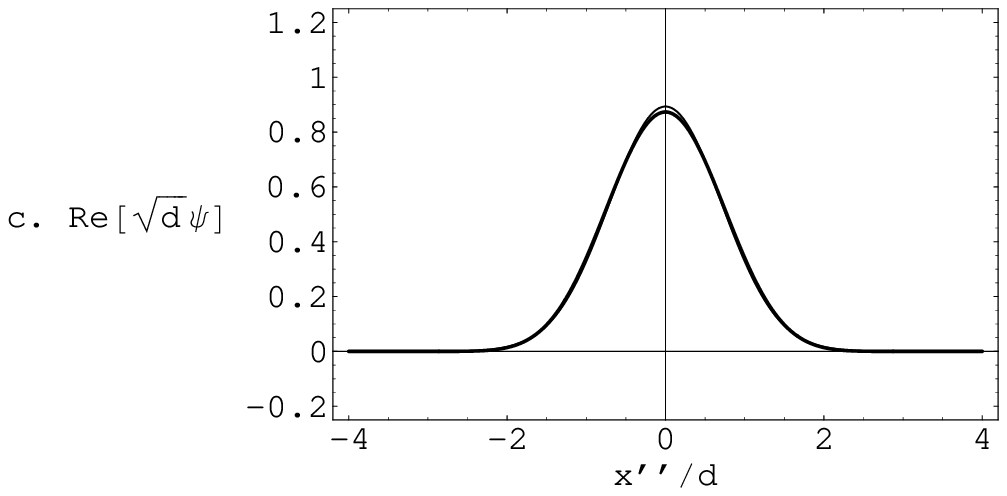}
			\epsfxsize=.93\linewidth
			\epsffile{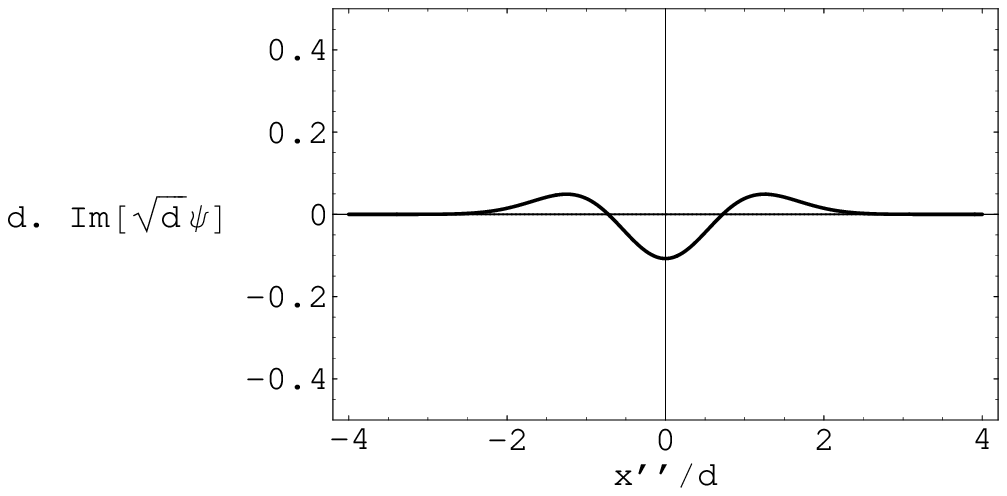}
	\end{center}
	\caption{Graph of $\sqrt{d} \psi$ 
	for the free particle where 
	$\psi$ equals $\psi(x'',0)$ (solid line), $\psi(x'',T)$ (small dashed 
	line), $\langle x'' | \hat{C}_{\Delta} | \psi \rangle$ (large 
	dashed line), and $\langle x'' | \hat{P}_{\Delta} | \psi \rangle$ (thick 
	line). For a. and b., $\delta = d$. For c. and d., $\delta = 10d$. In all graphs,
	$T = t_{\mathrm{spread}} / 4$.}
	\label{fig:fig1}
	\end{figure}

Our next task is to calculate $\langle x'' | \hat{C}_{\Delta} | x' 
\rangle$ according to Eq. \eqref{eq:classe}. For the free particle, the effective Lagrangian  
is Eq. \eqref{efflagrang}.
This is a well known 
system with a well known propagator \cite{FH65}:
	\begin{eqnarray}
	\int_u \delta x \, e^{iS_{\rm{eff}}[x(t)]/\hbar} &=&
	\left( \frac{m}{2 \pi i \hbar T} \right)^{1/2}
	\exp{ \left\{ \frac{i}{\hbar} \left[ \frac{m(x''-x')^{2}}{2T} \right. \right. }
	\nonumber
	\\
	&+&
	\frac{1}{2} \left( \frac{\hbar k}{T} \right) T(x''+x') \nonumber \\ &-&
	\left( \frac{\hbar k}{T} \right)^{2}
	\left. \left. \frac{T^{3}}{24m} \right] \right\}\, .
	\end{eqnarray}
Substituting this expression into Eq.~(\ref{eq:classeff}) and evaluating 
the integral over $k$ yields
	\begin{equation}
	\langle x'' | \hat{C}_{\Delta} | x' \rangle
	= K(x'',T; x', 0) E_\Delta   
	\left(\frac{x''+x'}{2}, \frac{\lambda}{\sqrt{3}} \right)
	\label{eq:eqforC}
	\end{equation}
where $\lambda$ was defined in Eq. \eqref{lambda}. 

Clearly the two operators Eq.~(\ref{eq:eqforP}) and Eq.~(\ref{eq:eqforC}) 
have a similar mathematical structure. Indeed they differ only by the
factor of $\sqrt{3}$ in the argument of the funtion $E_\Delta$. However, they 
are not equal. Examining how $\hat{C}_{\Delta}$ and $\hat{P}_{\Delta}$ act on a 
known wave function will give us an idea of how 
the two operators differ. To that end we examine $\langle x'' | 
\hat{C}_{\Delta} | \psi \rangle$ and $\langle x'' |
\hat{P}_{\Delta} | \psi \rangle$ where $| \psi \rangle$ has a Gaussian 
wave function of width $d$. 
	\begin{equation}
		\psi(x) = \left( \frac{2}{\pi d^2}\right)^{1/4} \exp{\left(-\frac{x^2}{d^2} \right)}
	\label{eq:gaussian}
	\end{equation}

With this Gaussian initial wave packet, we might expect classical behavior 
for spacetime alternatives when two conditions are satisfied: (1) The 
coarse-graining of position $\delta$ is much larger than the quantum 
uncertainty in position $d$ specified by the wave packet.  (2) The time $T$ over which the alternative 
is defined is much smaller than the wave packet spreading time of order the combination
\begin{equation}
t_{\rm spread} \equiv d^2 m / 2 \hbar  \  .
\label{spread}
\end{equation}
These two conditions can be neatly
expressed in terms of the length $\lambda$ introduced in
Eq.\eqref{lambda} as
	\begin{equation}
	\frac{\delta}{\lambda} \gg \frac{d}{\lambda} \gg 1 \ .
	\label{ineq}
	\end{equation} 
This is also the condition under which $E_\Delta$ is well approximated
by a top-hat function as discussed earlier.  Eq.\eqref{lambda} shows that typical
``macroscopic'' coarse grainings satisfy these criteria easily. 

Fig. \ref{fig:fig1} shows the real and imaginary parts of 
$\langle x'' | \hat{C}_{\Delta} | \psi \rangle$ 
and $\langle x'' | \hat{P}_{\Delta} | \psi \rangle$ for two cases, each
with $T= t_{\rm spread} / 4$,  but with $\delta = d$ in one case 
and $\delta = 10 d$ in the other. As expected, 
the two quantities are significantly different in the first case
when Eq.\eqref{ineq} is not satisfied,  and very close in the second 
case when it is satisfied.  
However, as discussed in Section II, for $\delta \gg d$, there is surprisingly good agreement between the time evolved wavefunction $\psi (x'',T)$, $\langle x'' | \hat{C}_{\Delta} | \psi \rangle$, and
$\langle x'' | \hat{P}_{\Delta} | \psi \rangle$, as can be seen in Fig. (\ref{fig:fig1}c-d.).

\subsection{The Classical Limit}
We now
compare $\langle x'' | \hat{C}_{\Delta} | x' \rangle$ and $\langle x'' |
\hat{P}_{\Delta} | x' \rangle$ in the classical limit, $\hbar 
\rightarrow 0$. Here we are dealing with a familiar 
situation in quantum mechanics. We have two distinct 
quantum operators, $\hat{C}_{\Delta}$ and $\hat{P}_{\Delta}$, each of 
which represents the same classical spacetime alternative. It is worth  
exploring whether the two operators represent exactly the same classical 
alternative.

For both $\hat{C}_{\Delta}$ and $\hat{P}_{\Delta}$, we use the 
method of stationary phase to evaluate 
the expressions in the $\hbar \rightarrow 0$ limit. Let us begin with 
$\hat{C}_{\Delta}$. Recall,
	\begin{equation}
	\langle x'' | \hat{C}_{\Delta} | x' \rangle = \int_u \delta x \,
	e_{\Delta}(\overline{x}[x(t)]) \, e^{iS[x(t)]/\hbar}\, .
	\label{eq:classe2}
	\end{equation}
We may introduce a change in path variable
	\begin{equation}
	x(t)=x_{\rm{cl}}(t)+y(t)
	\label{eq:xcl}
	\end{equation}
where $x_{\rm{cl}}(t)$ is the classical path. That is, $x_{\rm{cl}}(t)$ 
satisfies the classical equations of motion for the action $S$ with  
$x_{\rm{cl}}(0)=x'$ and $x_{\rm{cl}}(T)=x''$. 
Substituting Eq.~(\ref{eq:xcl}) into Eq.~(\ref{eq:classe2}) above gives:
	\begin{align}
	\langle x'' | \hat{C}_{\Delta} | x' \rangle &\approx 
	\int _u\delta y \, e_{\Delta}(\overline{x}[x_{\rm{cl}}(t) + 
	{y}(t)]) \, e^{iS[x_{\rm{cl}}(t)]/\hbar} \nonumber \\ &\times
	\exp{\left( \frac{i}{\hbar} \int_{0}^{T} dt \, \frac{1}{2}m
	\dot{y}^2 \right) }\, .
	\end{align}
where we have used $L=m\dot{x}^{2} /2$ and assumed there is
a unique classical path. In the 
$\hbar \rightarrow 0$ limit, the integral is dominated by the value 
of the integrand at the saddle point $y=0$. Thus, the top hat
function can be pulled out  of the path integral. Noting that
the remaining path integral 
is the propagator for the free particle with $y'=y''=0$ we find,
	\begin{equation}
	\langle x'' | \hat{C}_{\Delta} | x' \rangle \sim
	e_{\Delta}(\overline{x}_{\rm{cl}})
	K(x'',T;x',0)
	\label{eq:classapprox}
	\end{equation}
where ${\bar x}_{\rm{cl}}$ is the time average \eqref{eq:ave2} of the classical path 
${ x}_{\rm{cl}}(t)$.
The classical equation of motion is $m \ddot{x}_{\rm{cl}}=0$. Solving this 
equation with the above boundary conditions yields a unique
classical path:
	\begin{equation}
	x_{\rm{cl}}(t)=(x''-x')\frac{t}{T}+x'\, .
	\end{equation}
Thus,
	\begin{equation}
	\overline{x}_{\rm{cl}}=\frac{x''+x'}{2}\, .
	\end{equation}
Substituting this  into Eq.~(\ref{eq:classapprox}) gives the following 
asymptotic form of the class operator for the free particle as $\hbar\to 0$:
	\begin{equation} 
	\langle x'' | \hat{C}_{\Delta} | x' \rangle \sim
	e_\Delta\left(\frac{x'+x''}{2}\right) K(x'',T;x',0)  \ .
	\label{foureighteen}
	\end{equation}

We evaluate the expression \eqref{eq:pop}  for $\langle x'' | \hat{P}_{\Delta} | x' 
\rangle$ in the $\hbar \rightarrow 0$ limit by careful application of the 
stationary phase approximation  \cite{BO78} 
	\begin{equation}
	\int_{a}^{b} dt \, e^{izh(t)}g(t)
	\sim
	\frac{e^{izh(t_{0})} \sqrt{2 \pi}}{\sqrt{z} \sqrt{\pm h''(t_{0})}}
	e^{\pm \pi i/4} g(t_{0})\, 
	\label{eq:stphthm}
	\end{equation}
as $z \rightarrow \infty$ where $h'(t_{0}) = 0$ for $t_{0} \in (a,\,b)$, and $h''(t_{0}) \neq 0$. We 
use the plus signs if $h''(t_{0}) > 0$, and the minus signs if 
$h''(t_{0}) < 0$.
Substituting  Eq.~(\ref{eq:xbarpos}) and Eq.~(\ref{eq:freeprop}) into Eq. \eqref{eq:pop}, and 
using Eq.~(\ref{eq:stphthm}) to evaluate the integrals over $y$ and 
$\overline{x}$ in the $\hbar \rightarrow 0$ limit yields
	\begin{equation}
	\langle x''| \hat{P}_{\Delta} | x' \rangle \sim
	e_\Delta\left(\frac{x'+x''}{2}\right) K(x'',T;x',0) .
	\label{fourtwenty}
	\end{equation}
The same result could be obtained from the observation made earlier that 
$E_\Delta(z,\ell)$ approaches $e_\Delta(z)$ as $\delta/\ell$ becomes large.

A comparison of  \eqref{foureighteen} and \eqref{fourtwenty} shows that the operators $\hat{C}_{\Delta}$ 
and $\hat{P}_{\Delta}$ coincide in the classical limit. Their classical predictions will be the same.

\section{The Harmonic Oscillator}
\setcounter{equation}{0}
\label{ho}

The calculation of $\hat{P}_{\Delta}$ and $\hat{C}_{\Delta}$ for the 
harmonic oscillator parallels that of the free particle. We begin with the 
harmonic oscillator Hamiltonian:
	\begin{equation}
	H=\frac{p^{2}}{2m}+\frac{1}{2}m \omega^{2} x^{2}\, .
	\end{equation}
Solving the Heisenberg equations for $x$ and $p$ and calculating 
$\overline{x}$ yields
	\begin{equation}
	\overline{x} = \frac{\sin{\omega T}}{\omega T} x_{0} +
	\frac{1 - \cos{\omega T}}{(\omega T)^2} \frac{p_{0}}{m} T \, .
	\end{equation}
Again we find the eigenstates of $\overline{x}$, and impose delta function 
normalization. In the position basis, $|\overline{x}\rangle$ is given by
\begin{widetext}
	\begin{equation}
	\langle x | \overline{x} \rangle =
	\sqrt{\frac{m}{\pi \hbar T}}
	\frac{\omega T / 2}{\left| \sin{(\omega T / 2)} \right| }
	\exp{\left[ \frac{i}{\hbar} \frac{m}{T} \frac{(\omega T)^{2}}{1 - \cos(\omega T)}
	\left( \overline{x}x-\frac{1}{2}\frac{\sin{\omega T}}{\omega T}x^{2}
	\right) \right]}\, .
	\end{equation}
\end{widetext}
Substituting this expression into Eq.~(\ref{eq:pop}) along with the 
propagator for the harmonic oscillator (cf.\cite{FH65}), and integrating over 
$y$ and $\overline{x}$ yields:
	\begin{equation}
	\langle x'' | \hat{P}_{\Delta} | x' \rangle
	= K(x'',T;x',0)
	E_\Delta\left(\overline{x}_{\rm{cl}}, \lambda_{\mathrm{P}} \right) \, ,
	\label{eq:eqforPho}
	\end{equation}
where $\overline{x}_{\rm{cl}}$ is the average \eqref{eq:ave2} of the
position along the classical path given by 
	\begin{equation}
	\overline{x}_{\rm{cl}} =
	\left( \frac{1-\cos{\omega T}}{\omega T \sin{\omega T}} \right) (x' + x'') \, .
	\label{eq:xclassical}
	\end{equation}
and $\lambda_P$ is defined by 
	\begin{equation}
	\lambda_{\mathrm{P}} \equiv \left[ \frac{2 \hbar T}{m} \frac{\omega T}{\sin{\omega T}}
	\left(\frac{1 - \cos{\omega T}}{(\omega T)^2} \right)^2 \right]^{1/2}
	\end{equation}

In order to calculate $\langle x'' | \hat{C}_{\Delta} | x' \rangle$ for 
the harmonic oscillator, we note that 
\begin{equation}
L_{\rm{eff}} = \frac{1}{2}m\dot{x}^{2} - 
\frac{1}{2}m \omega^{2} x^{2} + \frac{\hbar k}{T}x \ .
\end{equation}
 Again, this system 
has a known propagator \cite{FH65}:
	\begin{equation}
	\int_u \delta x \, e^{iS_{\rm{eff}}/\hbar} = 
	\sqrt{ \frac{m \omega}{2 \pi i \hbar \sin{ \omega T}}}
	e^{iS_{\rm{eff,cl}}/\hbar}
	\label{eq:seffho}
	\end{equation}
where
	\begin{eqnarray}
	S_{\rm{eff,cl}} &=& 
	\frac{m \omega}{2 \sin{ \omega T}} \left. \bigg\{ (x^{\prime\prime2}+x^{\prime 2})
	\cos{ \omega T} - 2x'x''
	\right.
	\nonumber
	\\
	&+& 
	\frac{2}{m \omega} \frac{\hbar k}{T}
	\frac{1-\cos{\omega T}}{\omega}(x'+x'')
	\\
	&-&
	\left.	
	\frac{2}{m^{2} \omega^{2}} \left( \frac{\hbar k}{T} \right)^{2}
	\left[ \frac{1-\cos{\omega T}}{\omega^{2}} - \frac{T}{2 \omega}
	\sin{\omega T} \right ]
	\right\}\, . \nonumber
	\end{eqnarray}
Substituting Eq.~(\ref{eq:seffho}) into Eq.~(\ref{eq:classeff}), and 
evaluating the integral over $k$ yields
	\begin{equation}
	\langle x'' | \hat{C}_{\Delta} | x' \rangle
	= K(x'',T;x',0)
	E_\Delta\left(\overline{x}_{\rm{cl}}, \lambda_{\mathrm{C}} \right) \, ,
	\label{eq:eqforCho}
	\end{equation}
where
	\begin{equation}
	\lambda_{\mathrm{C}} \equiv \left[ \frac{4 \hbar T}{m} \frac{\omega T}{\sin{\omega T}}
	\left(\frac{1 - \cos{\omega T}}{(\omega T)^4} - 
        \frac{1}{2} \frac{\sin{\omega T}}{(\omega T)^3}
	\right) \right]^{1/2}
	\end{equation}
and $\overline{x}_{\rm{cl}}$ is defined in  
Eq.~(\ref{eq:xclassical}).

\begin{figure}[t!]
	\begin{center}
			\epsfxsize=.93\linewidth
			\epsffile{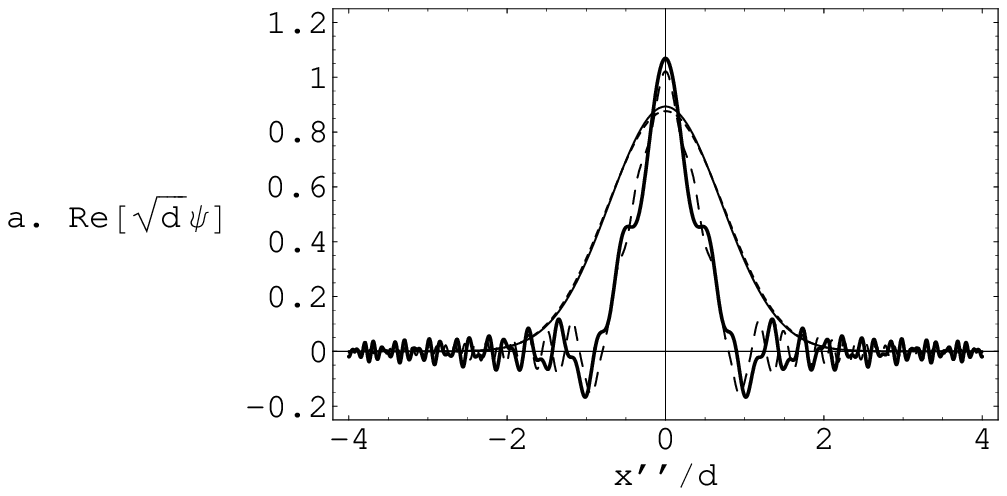}
			\epsfxsize=.93\linewidth
			\epsffile{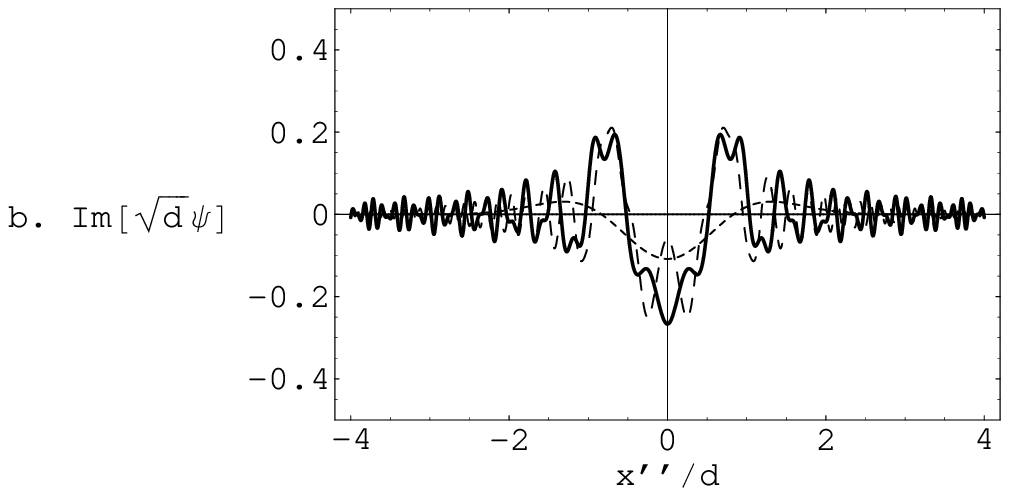}
			\epsfxsize=.93\linewidth
			\epsffile{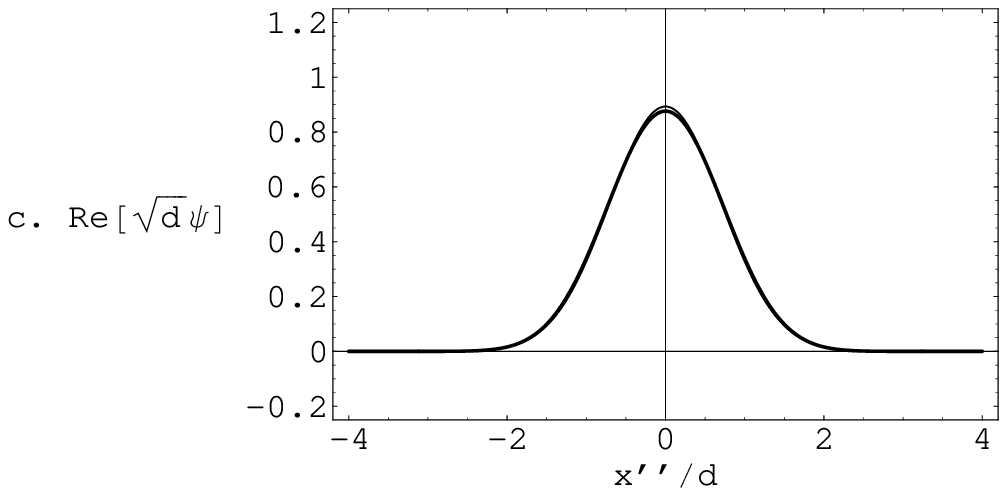}
			\epsfxsize=.93\linewidth
			\epsffile{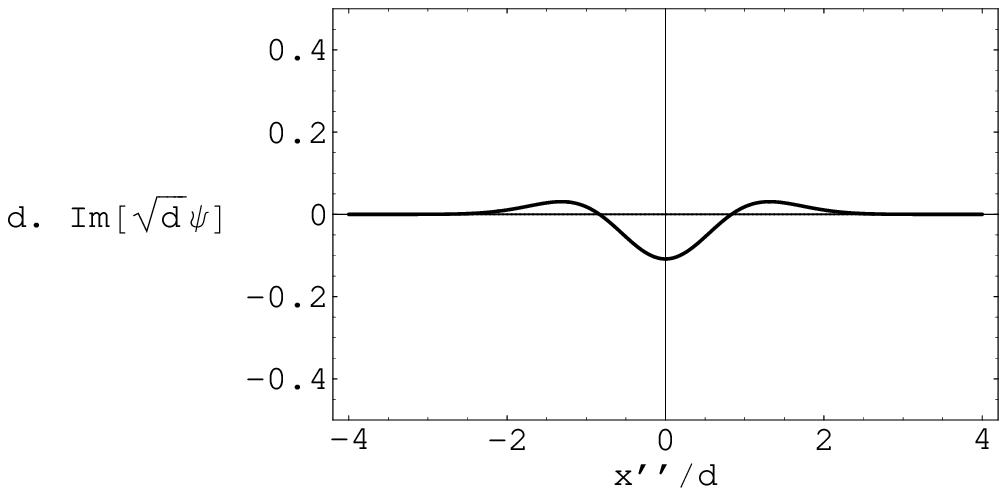}
	\end{center}
	\caption{Graph of $\sqrt{d} \psi$
	for the harmonic oscillator where
	$\psi$ equals $\psi(x'',0)$ (solid line), $\psi(x'',T)$ (small dashed
	line), $\langle x'' | \hat{C}_{\Delta} | \psi \rangle$ (large
	dashed line), and $\langle x'' | \hat{P}_{\Delta} | \psi \rangle$ (thick
	line). For a. and b., $\delta = d$. For c. and d., $\delta = 10d$. In 
        all graphs,
	$T = t_{\mathrm{spread}} / 4$ where $t_{\rm{spread}}$ is given by 
        Eq. \eqref{spread}.  
	}
	\label{fig:fig2}
	\end{figure}

Clearly $\hat{C}_{\Delta}$ and $\hat{P}_{\Delta}$ share a similar 
mathematical structure. However, they are not equal. Again, we examine how 
$\hat{C}_{\Delta}$ and $\hat{P}_{\Delta}$ act on a Gaussian 
wave function. The comparison parallels that of the free particle in Section 
IV. That is, for $\delta \approx d$, $\langle
x'' |\hat{C}_{\Delta} | \psi \rangle$ and $\langle x'' | \hat{P}_{\Delta}
| \psi \rangle$ look rather different (Fig. (\ref{fig:fig2}a-b.)). 
However, in 
Fig. (\ref{fig:fig2}c-d.), we see that for $\delta \gg d$, there is 
surprisingly good agreement between $\langle x'' | \hat{C}_{\Delta} | \psi 
\rangle$ and $\langle x'' | \hat{P}_{\Delta} | \psi \rangle$. Thus, we 
expect the two operators to produce similar results in classical 
situations.

We examine the classical limit.  The integrals in the $\hbar \rightarrow 
0$ limit can be evaluated using the same techniques as in Section IV. As
in the case of the free particle the matrix elements of the 
two operators are equal in the classical limit:
	\begin{equation}
	\langle x'' | \hat{C}_{\Delta} | x' \rangle \approx
	\langle x'' | \hat{P}_{\Delta} | x' \rangle \sim
	e_{\Delta}(\overline{x}_{\rm{cl}})
	K(x'',T;x',0)
	\end{equation}
where ${\bar x}_{\rm{cl}}$ is given by Eq. \eqref{eq:xclassical}.  
All the relations for the free particle discussed in the previous
section can be recovered from the zero frequency  limit of the relations
for the harmonic oscillator in this section.

\section{General Alternatives and General Potential}
\label{genalt}
The preceding sections demonstrate by example  the agreement, in a formal 
$\hbar\to 0$ classical  limit, between probabilities
calculated using class
operators and Heisenberg picture projection operators for ranges of
the spacetime alternative $\bar x$. This
section sketches a demonstration of this classical agreement for
general spacetime alternatives defined in terms of position and for general
Hamiltonians of the form Eq.~\eqref{threeone}.   Our results are essentially
formal and not rigorous but suggest  the underlying mechanisms of classical agreement.

We continue with a particle moving in one dimension described by a
Hamiltonian of the form Eq.~\eqref{threeone}.  The alternatives of interest are
specified classically by a functional $F_{\rm{cl}} [x(t)]$ of the particle
path $x(t)$ between $t=0$ and $t=T$. For example, the spacetime alternative
that the position time average $\bar x$ lies in a range $\Delta$
corresponds to the functional $F_{\rm{cl}}[x(t)]=e_\Delta (\bar x[x(t)])$ where $\bar x[x(t)]$ 
is the functional defined by Eq.~\eqref{threefour}. 
These are the alternatives for
which matrix elements of class operators can be defined by
sums-over-histories of the form
	\begin{equation}
	\langle x^{\prime\prime} |\hat F_{soh}|
	\, x^\prime \rangle = \int_{u}
	\delta x\ F_{\rm{cl}} [x(t)]\ e^{iS[x(t)]/\hbar}
	\label{sixone}
	\end{equation}
where the integration is over all paths $x(t)$ which start at $x^\prime$ at
$t=0$ and end at $x^{\prime\prime}$ at $t=T$. The sum over histories
Eq.~\eqref{onethree} is of this form with $F_{\rm{cl}} [x(t)] = e_\Delta (\bar x[x(t)])$.

The formal classical limit of path integrals like Eq.~\eqref{sixone} is
easily calculated. Assuming
that $F_{\rm{cl}}$ does not itself depend on $\hbar$, the dominant contribution as $\hbar \to 0$ comes from
classical paths $x_{\rm{cl}}(t)$ between $x'$ at $t=0$ and $x''$ at $t=T$ that extremize the action $S[x(t)]$. Assume for simplicity
that there is only one such classical path $x_{\rm{cl}} (t, x^{\prime\prime},
x^\prime)$. Because of the increasingly rapid varying phase as $\hbar \to 0$,
the functional $F_{\rm{cl}}$ may be taken outside the integral and evaluated at
this $x_{\rm{cl}}(t, x^{\prime\prime}, x^\prime)$. Thus,
	\begin{align}
	\langle x^{\prime\prime}|\hat F_{soh}|
	x^\prime\rangle &= F_{\rm{cl}}\left[x_{\rm{cl}}
	\left(t, x^{\prime\prime}, x^\prime\right)\right]
	\, K \left(x^{\prime\prime}, T\, ;\, x^\prime, 0\right) \nonumber \\
	&+ \epsilon(x'',x') \ ,
	\label{sixtwo}
	\end{align}
where  $K(x^{\prime\prime}, T\, ;\, x^\prime, 0)$ is the propagator defined
by the unrestricted, unweighted, sum-over-paths [\cf Eq.~\eqref{twoseven}] . Here and throughout,  {\it by $\epsilon(x'',x')$ we mean  some function that goes to zero with $\hbar$}, typically like
$\sqrt{\hbar}$.

We now demonstrate the same classical limit for the matrix elements of 
\begin{equation}
\hat F_{Hp} \equiv e^{-iHT/\hbar} F
\label{sixthree}
\end{equation}
where $F$ is a Hermitian Heisenberg picture operator representing the
classical functional $F_{\rm{cl}} [x(t)]$. Different operator representations
can be formed by first  using the classical equations of motion to express
$F_{\rm{cl}}$ as functions of the position $x_0$ and momentum $p_0$ at time
$t=0$. (Eq.~\eqref{fourone} is an explicit example for the time average
$\bar x$. ) Then, the different Hermitian operator orderings for this classical
expression give different Heisenberg picture operators $F$ representing the
classical functional $F_{\rm{cl}}$. We expect matrix elements of these to
agree in the classical limit and they will.  

Our construction of the classical limit of Eq.~\eqref{sixthree} relies on
extrapolating results from microlocal analysis \cite{Omn94}. The central
element of this formalism is the {\it symbol} $\check{A} \, (X,
P)$ of an operator $A$ defined by 
	\begin{subequations}
	\label{sixfour}
	\begin{equation}
	\check{A}\, (X, P)  =  \int\, d\xi\, e^{-iP\xi/\hbar}
	\langle X + \xi/2 | A|X-\xi/2\rangle\, , 
	\label{sixfour a}\\
	\end{equation}
with the inverse formula
	\begin{align}
	\left\langle x^{\prime\prime}|A|x^\prime\right\rangle  =&  \frac{1}{2\pi}
	\int dX \int dP\, \delta\left(X - \frac{x^{\prime\prime}
	+ x^\prime}{2}\right) \nonumber \\ &\times  e^{iP(x^{\prime\prime}-x^\prime)/\hbar}
	\ \check{A}\, (X,P)\, .
	\label{sixfour b}
	\end{align}
	\end{subequations}
In the simple case $A=|\psi\rangle\langle\psi|$, 
$\check{A}\, (X,P)$ is the Wigner distribution for the state $|\psi\rangle$.

Two results from microlocal analysis concerning the classical limit
will be important for us. The first concerns the symbol for the product of
two operators. If $C=AB$ then
	\begin{equation}
	\check{C}\, (X,P) =
	\check{A}\, (X,P)\, \check{B}\, (X,P)
	+ \epsilon(X,P) \, .
	\label{sixfive}
	\end{equation}
The second result concerns the time evolution of the operator $A(t)$ given by the
Heisenberg equations of motion Eq.~\eqref{threetwo}. Express $A(t)$ in terms of
the position and momentum operators $x_0$ and $p_0$ at $t=0$. The relation
$x(t)= x_0 + p_0 t/m$ for a free particle is a simple example.  Construct
the symbol $ \check{A}\, (t, X,P)$ using eigenstates of $x_0$ as
the basis in Eq.~\eqref{sixfour}. Construct the symbol of the
$\check{H}\, (X,P)$ of the
Hamiltonian in the same way. Then, to leading order in $\hbar$, the symbol
for $A$ obeys the classical equation of motion. 
	\begin{equation}
	\frac{\partial \check{A}}{\partial t} = 
	\{\check{A},\, \check{H}\}
	+ \epsilon(X,P) 
	\label{sixsix}
	\end{equation}
where $\{\cdot\, ,\, \cdot\}$ is the Poisson bracket. Explicit forms of 
the corrections to Eq.~\eqref{sixfive} and Eq.~\eqref{sixsix} are given in
\cite{Omn94}. We now employ these
two results to calculate the classical limit of Eq.~\eqref{sixthree}.

Matrix elements of Eq.~\eqref{sixthree} in the basis of eigenstates of $x_0$
can be written
	\begin{equation}
	\langle x^{\prime\prime}|\hat F_{Hp}|x^\prime\rangle
	= \int dy\, \langle x^{\prime\prime}|e^{-iHT/\hbar}|
	y\rangle\, \langle y|F|x^\prime\rangle\, .
	\label{sixseven}
	\end{equation}
We concentrate first on the classical limit of $\langle
x^{\prime\prime}|F|x^\prime\rangle$ and later return to that for the whole
expression Eq.~\eqref{sixseven}.  

We begin with the time evolution of the symbols for $x$ and $p$. At $t=0$, 
$\check{x}(0) = X$ and $\check{p}(0)=P$. Their
time evolution to leading order in $\hbar$ is determined by the two coupled
classical equations \eqref{sixsix} for $\check{x} (t)$ and
$\check{p} (t)$ --- Hamilton's equations of motion. We can
therefore write
	\begin{equation}
	\check{x}\, (t, X, P) = x_{\rm{cl}} (t, X, P) + \epsilon(t,X,P)
	\label{sixeight}
	\end{equation} 
where $x_{\rm{cl}}(t, X, P)$ is the solution of the classical equation of
motion with initial $(t=0)$ position $X$ and momentum $P$. For example, the
symbol for time average $\bar x$ is
	\begin{equation}
	\check{\bar{x}} (X, P) = \bar x_{\rm{cl}}(X, P) + \epsilon(X,P)
	\label{sixnine}
	\end{equation}
where $\bar x_{\rm{cl}}$ is the classical time average. Some limitation on the
size of the time $T$ is likely to be needed to control the size of the corrections $\epsilon$. 

More general  functionals of $x(t)$ can be treated as follows. First,
solve the Heisenberg equations of motion and express $x(t)$ in terms of
operator products of $x_0$'s and $p_0$'s.  Use Eq.~\eqref{sixfive} to show that
the symbols of these products are the product of the symbols to leading
order $\hbar$. Therefore, for any operator representative $F$ of the classical
functional $F_{\rm{cl}}[x(t)]$, the classical limit of its symbol is
	\begin{equation}
	\check{F}\, (X,P) = F_{\rm{cl}}[x_{\rm{cl}}(t, X, P)] 
	+ \epsilon(X,P) \, .
	\label{sixten}
	\end{equation}
We now use this formula to calculate the classical limit of the matrix
elements of $\langle x^{\prime\prime}|\hat F_{Hp}|x^\prime\rangle$ defined
using Eq.~\eqref{sixfour}.

Using Eq.~\eqref{sixfour b}, the expression Eq.~\eqref{sixseven} can be written
	\begin{eqnarray}
	\langle x^{\prime\prime}|\hat F_{Hp}|
	x^\prime\rangle & = & \frac{1}{2\pi}\ \int dy \int dP
	\, \langle x^{\prime\prime}|e^{-iHT/\hbar}|y\rangle
	\nonumber\\
	& \times & e^{iP(y-x^\prime)/\hbar} \check{F}
	\left(\frac{y+x^\prime}{2}\, ,\, P\right)\, .
	\label{sixeleven}
	\end{eqnarray} 
The classical limit of the symbol $\check{F}$ is given by
Eq.~\eqref{sixten}. The classical limit of the propagator is given by the
standard result
	\begin{align}
	&\langle  x^{\prime\prime}|e^{-iHT/\hbar}|y\rangle 
        \equiv
	K\left(x^{\prime\prime}, T\, ;\, x^\prime 0\right) \nonumber \\
	&\quad = D^{-\frac{1}{2}}\left(x^{\prime\prime}, y\right)\, \exp
	\, \left[iS_{\rm{cl}}\left(x^{\prime\prime}, y, T\right)/\hbar\right] + \epsilon(x'',y) \, .
	\label{sixtwelve}
	\end{align}
Here, $S_{\rm{cl}}(x^{\prime\prime}, y, T)$ is the action of the classical
path $x_{\rm{cl}}\, (t, x^{\prime\prime}, y)$
(assumed unique) that starts at $y$ at $t=0$ and ends at
$x^{\prime\prime}$ at $t=T$. The slowly varying prefactor $D$ is essentially
the DeWitt-van Vleck determinant. Inserting Eq.~\eqref{sixtwelve} and
Eq.~\eqref{sixten} into Eq.~\eqref{sixeleven}, we find
\begin{widetext}
	\begin{equation}
	\langle x^{\prime\prime}|\hat F_{Hp}|x^\prime\rangle
	 =  \frac{1}{2\pi}\ \int dy \int dP\, D^{-\frac{1}{2}}
	\left(x^{\prime\prime}, 
	y\right)\ F_{\rm{cl}}\left[x_{\rm{cl}}\left(t, \frac{y+x^\prime}{2}\, ,
	\, P\right)\right]
	  \exp\, \left\{\frac{i}{\hbar}\
	\left[S_{\rm{cl}}\left(x^{\prime\prime}, y, T\right) +
	P\left(y-x^\prime\right)\right]\right\} + \epsilon(x'',x') \, .
	\label{sixthirteen}
        \end{equation}
\end{widetext}

The rapid variation of the phase for small $\hbar$ means that the double
integral in Eq.~\eqref{sixthirteen} can be evaluated by the method of 
stationary phase. The dominant
contribution comes from the $(y, P)$ that extremize the exponential's
phase.  The conditions for this are
	\begin{equation}
	y=x^\prime\quad , \quad P=-\frac{\partial S_{\rm{cl}}}{\partial y}\, .
	\label{sixfourteen}
	\end{equation}
The second of these singles out the initial momentum $P(x^{\prime\prime},
x^\prime, T)$ necessary for the classical path to arrive at
$x^{\prime\prime}$ a time $T$ after it starts at $x^\prime$. We thus find
for the leading contribution
	\begin{align}
	\langle x^{\prime\prime}|\hat F_{Hp}|x^\prime\rangle
	&= F_{\rm{cl}} \left[x_{\rm{cl}} \left(t, x^{\prime\prime}, x^\prime\right)\right]
	\ K\, (x^{\prime\prime}, T\, ;\, x^\prime, 0) \nonumber \\
       & + \epsilon(x'',x') 
	\label{sixfifteen}
	\end{align}
where $x_{\rm{cl}}(t, x^{\prime\prime}, x^\prime)$ is the classical path from
$x^\prime$ to $x^{\prime\prime}$ in time $T$. 

Comparison of Eq.~\eqref{sixfifteen} with Eq.~\eqref{sixtwo} shows that the matrix
elements of $\langle x^{\prime\prime} |\hat F_{Hp}|
x^\prime\rangle$ agree with $\langle
x^{\prime\prime}|\hat F_{soh}|x^\prime\rangle$ in
leading classical order.  This is the agreement demonstrated 
explicitly in Sections IV and V for the
time average $\bar x$ for the free particle and harmonic oscillator. The
argument in this section is a general form of that.

\section{Decoherence of Spacetime Alternatives in the Classical Limit}
Previous sections have shown how the representations of spacetime alternatives
by sum-over-histories class operators and Heisenberg picture projection operators
coincide in the classical limit. A general argument was given in Section \ref{genalt}
and Sections \ref{freepart} and \ref{ho} provided specific examples. This section
demonstrates the decoherence of exhaustive sets of histories coarse-grained by
spacetime alternatives in the same classical limit. The essence of the argument is this:
Decoherence is automatic and exact for sets of alternatives represented by
projections [\cf Eq.~\eqref{orthproj}]. It is not necessarily automatic for sets represented by sum-over-histories class operators. However, in classical limit, where
the class operators coincide with projections, sum-over-histories class operators
must decohere also. 

We continue to consider the histories of a particle moving in one dimension with
a Hamiltonian Eq.~\eqref{threeone} between time $0$ and $T$. To keep the discussion
manageable we restrict attention to coarse grainings by ranges of values of a functional
of the paths $f[x(t)]$ .  The classical functionals $F_{\rm{cl}}[x(t)]$ considered in the previous section are a special case. The time average $\bar x$ of Eq.~\eqref{eq:ave2} is an example.
Specifically, we consider an exhaustive set of exclusive ranges $\{\Delta_\alpha\}$,
$\alpha = 1, 2, \cdots$ and the set of alternative histories defined by the  classes
$\{c_\alpha\}$ where $f[x(t)]$ takes values in these ranges. The class operators for the set are defined by [\cf Eq.~\eqref{onethree}, Eq.~\eqref{sixone}]  
\begin{equation}
	\langle x'' | \hat{C}_{\Delta} | x' \rangle \equiv 
	\int _u \delta x \, e_\Delta(f[x(t)]) 
	\exp{\left( \frac{i}{\hbar}S[x(t)] \right) }\, . 
	\label{sevenone}
	\end{equation}
where we denote by $\Delta$ any of the intervals in the set $\{\Delta_\alpha\}$.
This set of histories decoheres when the condition Eq.~\eqref{meddec} is satisfied.

Projection operators onto distinct ranges are exactly orthogonal [\cf Eq.~\eqref{orthproj}].
	\begin{equation}
	P_\Delta P_{\Delta'}  =\hat P^\dagger_\Delta \hat P_{\Delta'} = 0, \quad \Delta \ne \Delta' .
	\label{seventhree}
	\end{equation}
The decoherence condition Eq.~\eqref{meddec} is therefore exact if the $\hat C$'s are
replaced by $P$'s or $\hat P$'s.  Since matrix elements of $\hat C_\Delta$ coincide with
those of $\hat P_\Delta$ in the $\hbar \to 0$ limit, we expect
Eq.~\eqref{meddec} to be 
satisfied in that limit as a consequence. We now demonstrate this explicitly.

Consider 
	\begin{align}
	\langle x | \hat C^\dagger_\Delta \hat C_{\Delta'} |x' \rangle  
	&=  \int dy \langle y | \hat C_\Delta |x\rangle^*\langle y | \hat C_{\Delta'} |x'\rangle
	\nonumber \\
	&= 
	\int dy \int_u \delta x \int_u \delta x'
	e_\Delta(f[x(t)])e_{\Delta'}(f[x'(t')]) \nonumber \\ 
        &\times \exp\left\{
	\frac{i}{\hbar} \left( S[x'(t')] -S[x(t)]\right)\right\} .
	\label{sevenfour}
	\end{align}
The integrals are over paths $x(t)$ and $x'(t')$ that start at $x$ and $x'$  respectively at $t=0$ and
$t'=0$  and both end at $y$ at $t=t'=T$.  In the $\hbar \to 0$ limit the rapid oscillation of the 
exponential means that the integrals will be dominated by the stationary (classical) paths
$x_{\mathrm{cl}}(t,y,x)$ and $x_{\mathrm{cl}}(t,y,x')$ connecting $x$ and $x'$ to $y$. (We suppose unique stationary paths for simplicity.) We assume that, as a consequence, the top-hat  factors can be taken outside the path integrals and evaluated at the stationary paths. Thus in the
$\hbar \to 0$ limit we have 
\begin{widetext}
	\begin{equation}
	\langle x | \hat C^\dagger_\Delta \hat C_{\Delta'} |x' \rangle\sim  \int dy \,
	e_\Delta(f[x_{\mathrm{cl}}(t, y,x)])
	e_{\Delta'} (f[x_{\mathrm{cl}}(t, y,x')]) \exp\left\{ \frac{i}{\hbar}
	\left( S_{\mathrm{cl}}(y,x')  -S_{\mathrm{cl}}(y,x)\right)\right\} ,
	\label{sevenfive}
	\end{equation}
\end{widetext}
where $S_{\mathrm{cl}}(y,x)$ is the action evaluated at the classical path with endpoints $y$ and
$x$. 

 When $x=x'$ The exponent in Eq.~\eqref{sevenfive} vanishes but so
does the factor multiplying it when $\Delta \ne \Delta'$. The reason is that when $x=x'$ the arguments of the top-hat functions are the same but their ranges are exclusive. 
When $x\ne x'$ we assume that the derivative of the  exponent with respect to $y$ does
not vanish identically. The Riemann-Lebesgue lemma then shows that
Eq.~\eqref{sevenfive} vanishes as $\hbar\to 0$. Thus for all values of $x$
and $x'$ we confirm Eq.~\eqref{meddec} and the decoherence of the set of histories in the classical limit.

The case of the histories of a free particle partitioned by ranges of the time average postion $\bar x$ that was worked out in detail in Section \ref{freepart} provides a ready example. Using Eq.~\eqref{eq:eqforC} the matrix element in Eq.~\eqref{sevenfour} is explicitly
\begin{widetext}
	\begin{align}
	\langle x | \hat C^\dagger_\Delta \hat C_{\Delta'} |x' \rangle  = \int dy \langle y | \hat C_\Delta
	|x\rangle^*\langle y | \hat C_{\Delta'} |x'\rangle &= \left(\frac{m}{2\pi\hbar T}\right)
	\exp\left[\frac{i}{\hbar}\frac{m}{2T}(x'^2 -x^2)\right]\int dy \exp\left[ \frac{i}{\hbar}\frac{m}{T}(x-x')y
	\right] \nonumber \\ &\times E_\Delta \left(\frac{y+x}{2}, \frac{\lambda}{ \sqrt{3}} \right)
	E_{\Delta'}\left(\frac{y+x'}{2}, \frac{\lambda}{\sqrt{3}} \right).
	\label{sevensix}
	\end{align}
\end{widetext}
As discussed in Section \ref{freepart}, $E_\Delta(z, \lambda / \sqrt{3}) \to e_\Delta(z)$ in the
limit $\hbar \to 0$ and $\lambda$ becomes large. This allows $E$'s to be replaced
by $e$'s in Eq.~\eqref{sevensix} evaluated at the classical path. For a classical path moving between
$x$ and $y$  in a time $T$, $\bar x = (y+x)/2$. This shows the argument of the 
$e$'s in Eq.~\eqref{sevenfive} is just that of the argument of the $E$'s in Eq.~\eqref{sevensix}.
Thus the form Eq.~\eqref{sevenfive} is recovered explicitly from Eq.~\eqref{sevensix}. 

\section{Conclusion}

The lesson of both the special and general theories of relativity is that 
four-dimensional
spacetime is the most general arena for physics on scales well above the Planck
scale of quantum gravity. Correspondingly quantum theory is formulated most
generally in four dimensional form in terms of sets of histories of 
spacetime alternatives that are extended in time, their decoherence, and 
their probabilities. This paper 
has explored the classical ($\hbar\to 0$)
limit of quantum operators representing spacetime alternatives in the 
context of non-relativistic quantum theory. A given classical spacetime 
alternative may have many different representations in terms of quantum operators. We considered
two kinds:  (1) Class operators defined by sums over the classical
histories of  the alternative, and (2) projection operators on ranges of the 
Heisenberg picture
operators. In Sections VI  and VII we gave a general
argument based on microlocal analysis for why the class operators for a set of
exclusive alternatives decohere in the classical limit and why the predictions of both kinds of representations coincide in that limit.   Our results were formal because we did not provide
general estimates for the corrections to decoherence and the differences
in probabilities for small $\hbar$. However, we analyzed these corrections 
explicitly for a particular alternative in particular tractable model
systems in Sections III and IV. 
Specifically we considered  the average of position over a time interval  $\overline{x}$
as a simple spacetime alternative for the free particle and harmonic oscillator in one-dimension.
We showed by explicit calculation how class operators and projections have differing  matrix elements in general, but also how those coincide in the classical limit. These results show explicitly
how class operators and corresponding projection operators can be different operator representations  of the same classical spacetime alternative.

\acknowledgments
We thank R. Omn\`es for a critical reading of the manuscript. 
This work was supported in part by NSF grant PHY02-44764.  The work of
AB was supported by the Frank H. and Eva B. Buck Foundation.


\begin{thebibliography}{99}

\bibitem{Haa96} R.~Haag, {\sl Local Quantum Physics}, Springer, Berlin 
(1996).

\bibitem{Har95c} J.B.~Hartle, {\it Spacetime Quantum Mechanics and the
Quantum Mechanics of Spacetime} in {\sl
Gravitation and Quantizations}, Proceedings of the 1992 Les Houches
Summer School, edited by B. Julia and J. Zinn-Justin, Les Houches Summer
School Proceedings Vol. LVII (North Holland, Amsterdam,
1995); gr-qc/9304006. 

\bibitem{BR33}  N.~Bohr and L.~Rosenfeld, {\sl Det Kgl. Danske Vidensk.
Selskab Mat.-Fys. Medd.}, {\bf 12}, nr. 8, 1933. [translated as {\it The
Measurability of the Electromagnetic Field} in
J.A.~Wheeler and W.H.~Zurek, eds {\sl Quantum Theory of
Measurement}, Princeton University Press, Princeton (1983)].

\bibitem{BR50} N.~Bohr and L.~Rosenfeld, {\it Field and Charge
Measurements in Quantum Electrodynamics}, {\sl Phys.~Rev.} {\bf 78}, 794
(1950).
 
\bibitem{Fey48} R.P. Feynman, {\it Space-time Approach to Non-Relativistic
Quantum Mechanics}, {\sl Rev. Mod. Phys.} {\bf 20}, 267 (1948). 

\bibitem{DeW62} B.~DeWitt, {\it The Quantization of Geometry} in {\sl Gravitation an Introduction to Current
Research}, ed.~by L.~Witten, John Wiley, New York (1962).

\bibitem{Men93} M.B.~Menskii, {\sl Continuous Quantum Measurements and Path
Integrals}, IOP Publishing, Bristol (1993);  {\sl Quantum
Measurements and Decoherence: Models and Phenomenology}, Kluwer, Dordrecht
(2000).

\bibitem{Har91b} J.B.~Hartle, {\sl Spacetime Coarse Grainings in 
Non-Relativistic Quantum Mechanics},  {\sl Phys.~Rev.~D} {\bf 44}, 3173 (1991).

\bibitem{YT91a} N.~Yamada and S.~Takagi,{\it Quantum Mechanical
Probabilities on a General Spacetime Surface},  {\sl Prog.~Theor.~Phys.} 
{\bf 85},
985 (1991); {\it Quantum Mechanical Probabilities on a General Spacetime 
Surface II.},  {\sl ibid} {\bf 86}, 599 (1991); {\it Spacetime
Probabilities in Non-Relativistic Quantum Mechanics}, {\sl ibid} {\bf 87},
77 (1992).

\bibitem{Sor93} R. Sorkin, {\sl Impossible Measurements of Quantum Fields} 
in {\it Directions in Relativity, vol 2, Papers in Honor of Dieter Brill}, 
edited by B.-L. Hu and T. Jacobson (Cambidge University Press, Cambridge,1993).

\bibitem{Mar94} D. Marolf, {\it Models of Particle Detection in Regions
of Spacetime}, {\sl Phys. Rev. A}, {\bf 50}, 939, 1994; 
gr-qc/9307003
 
 
\bibitem{Halxx} J.~Halliwell, {\sl The Interpretation of Quantum Cosmology and
the Problem of Time}, in {\it The Future of Theoretical Physics and
Cosmology}, ed. by G.W.~Gibbons, E.P.S~Shellard, and S.J.~Rankin,
Cambridge University Press, Cambridge, 2003, gr-qc/0208018, and 
{\sl Decoherent Histories and Spacetime Domains}, in {\it Time and
Quantum Mechanics}, ed. by J.G.Muga, R. Sala Mayato and
I.L.Egusquiza, Springer, Berlin, 2001, quant-ph/0101099.

\bibitem{Har04} J. B. Hartle, {\it Linear Postivity and Virtual Probability}, {\sl Phys.~Rev.~A},
{\bf70}, 022104 (2004), quant-ph/0401108.


\bibitem{FH65} See, {\it e.g.}
R.P.~Feynman and A.~Hibbs, {\sl Quantum Mechanics and Path
Integrals}, McGraw-Hill, New York (1965).


\bibitem{BO78} See, {\it e.g.} J.E.~Marsden and M.J.~Hoffman, {\sl Basic Complex Analysis}, W.H.~Freeman, New York (1999).

\bibitem{Omn94} R.~Omn\`es, {\sl Interpretation of Quantum Mechanics},
Princeton University Press, Princeton (1994), {\it Quantum-Classical 
Correspondence Using Projection Operators}, {\sl J. Math. Phys.} {\bf 38}, 697 (1997). 


\end{thebibliography}
\end{document}